\documentclass[acmtog]{acmart}

\AtBeginDocument{%
  \providecommand\BibTeX{{%
    \normalfont B\kern-0.5em{\scshape i\kern-0.25em b}\kern-0.8em\TeX}}}

\setcopyright{acmcopyright}
\copyrightyear{2023}
\acmYear{2023}
\acmDOI{XXXXXXX.XXXXXXX}

\usepackage[english]{babel}

\usepackage{amsmath}
\usepackage{graphicx}
\usepackage{mdframed}
\usepackage{listings}
\usepackage{booktabs}
\usepackage{subcaption}

\acmJournal{TOG}
\acmVolume{37}
\acmNumber{4}
\acmArticle{111}
\acmMonth{8}

\begin{document}
\newcommand{\sender}{\mathsf{s}}
\newcommand{\recipient}{\mathsf{r}}
\newcommand{\owner}{\mathsf{o}}
\newcommand{\stealth}{\mathsf{st}}
\newcommand{\pubkey}{\mathsf{pk}}
\newcommand{\registrant}{\mathsf{reg}}

\newtheorem{heuristic}[theorem]{Heuristic}

\title[Anonymity Analysis of the Umbra Stealth Address Scheme on Ethereum]{Anonymity Analysis of the Umbra Stealth Address Scheme on Ethereum}


\author{Alex M. Kovács}
\authornote{}
\email{kovcsaleex0104@gmail.com}

\author{István András Seres}
\orcid{0000-0003-0143-4057}
\email{seresistvanandras@gmail.com}
\affiliation{%
  \institution{Eötvös Loránd University}
  \streetaddress{Pázmány Péter stny. 1/C}
  \city{Budapest}
  \state{Pest}
  \country{Hungary}
  \postcode{1117}
}


\renewcommand{\shortauthors}{Kovács and Seres}

\begin{abstract}
  Stealth addresses are a privacy-enhancing technology that provides recipient anonymity on blockchains. In this work, we investigate the recipient anonymity and unlinkability guarantees of Umbra, the most widely used implementation of the stealth address scheme on Ethereum and on its three off-chain scalability solutions, e.g., Arbitrum, Optimism, and Polygon. We define and evaluate four heuristics to uncover the real recipients of stealth payments. We find that for the majority of Umbra payments, it is straightforward to establish the recipient, hence nullifying the benefits of using Umbra. Specifically, we find the real recipient of $48.5\%$, $25.8\%$, $65.7\%$, and $52.6\%$ of all Umbra transactions on the Ethereum main net, Polygon, Arbitrum, and Optimism networks, respectively. Finally, we suggest easily implementable countermeasures to evade our deanonymization and linking attacks.
\end{abstract}



\received{20 February 2007}
\received[revised]{12 March 2009}
\received[accepted]{5 June 2009}

\maketitle

\section{Introduction}
Decentralized finance (DeFi) offers a radically new financial system, i.e., it is an open, trustless,  composable, and transparent financial system built on top of blockchains. The transparent nature of DeFi aids trustlessness and public verifiability. However, the lack of financial privacy is undesirable in most applications. Therefore, the DeFi ecosystem on Ethereum offers several deployed privacy-enhancing solutions, such as mixers (e.g., Tornado Cash~\cite{pertsev2019tornado}), confidential transactions (e.g., AZTEC~\cite{williamson2018aztec}) or stealth addresses (e.g., Umbra~\cite{difrancesco2021umbra}. Previous work has thoroughly analyzed the anonymity guarantees provided by Tornado Cash~\cite{beres2021blockchain,wang2023zero,wu2022tutela}. In this work, we study the recipient anonymity and unlinkability guarantees provided by the Umbra stealth address scheme. 

Stealth addresses (or seldom Elliptic-curve Diffie–Hellman (ECDH) addresses) are a prevalent privacy-enhancing technology for cryptocurrencies. They were first proposed by Peter Todd on the Bitcoin mailing list in 2014~\cite{todd2014stealth}. Stealth addresses allow a payment sender to transfer assets \emph{non-interactively} to a recipient in a recipient anonymous way. In a nutshell, senders generate a computationally random-looking address, i.e., a stealth address. Subsequently, the sender can pay the recipient by transferring funds to the freshly generated stealth address. Specifically, no party other than the transacting parties can establish the actual recipient of a transaction using stealth addresses. However, the recipient can detect on the blockchain that a specific stealth address belongs to them, and crucially they are the only ones who can spend the funds at the stealth address. In conclusion, stealth addresses allow the sender to obfuscate its economic relationship with the recipient.

Stealth addresses are standardized for Bitcoin in the Bitcoin Improvement Proposal (BIP) 47~\cite{ranvier2015bip47} and Ethereum in the Ethereum Improvement Proposal (EIP) 5564~\cite{wahrstatter2022eip5564}. Even though the BIP47 did not gain significant traction in the Bitcoin ecosystem~\cite{moser2017anonymous}, stealth addresses remain a popular privacy-enhancing technique for public blockchains as they are implemented and deployed in Monero~\cite{noether2014monero} and Ethereum~\cite{wood2014ethereum} as well.

This paper aims to analyze the recipient anonymity guarantees achieved by the most popular stealth address implementation on Ethereum: Umbra\footnote{See:~\url{https://app.umbra.cash/}.}. Umbra is currently deployed on Ethereum and its numerous layer-2 scalability solutions such as Polygon, Arbitrum, and Optimism. Umbra is the second most popular privacy-enhancing technology used on Ethereum and its layer-2 systems after the AZTEC protocol with $413$ daily users versus $1866$ daily users on March 20th, 2023\footnote{See:~\url{https://dune.com/intake/umbra-protocol}}\textsuperscript{,}\footnote{See:~\url{https://dune.com/gm365/aztec-v2}.}. In this work, our goal is to identify erroneous idioms of use that allow any blockchain analysts to reduce the theoretical recipient anonymity and unlinkability guarantees provided by the stealth address cryptographic scheme.

In this work, we make the following contributions.
\begin{itemize}
    \item We identify four heuristics enabled by user behavior that can reduce or nullify the recipient anonymity or unlinkability guarantees of the Umbra stealth address scheme, see Section~\ref{sec:heuristics}.
    \item We evalaute the efficacy of the identified heuristics in uncovering and linking Umbra stealth payment recipients, see Section~\ref{sec:evaluation}. Our results are reproducible as our code is open-source. It is available at the following repository:~\url{https://github.com/alekszkovacs/UmbraAnonymityAnalysis}.
    \item We suggest counter-measures to our proposed heuristics. We make suggestions to stealth wallet providers and developers.  
\end{itemize}

The rest of this paper is organized as follows. In Section~\ref{sec:background}, we describe the cryptographic details of stealth addresses and, in particular, the relevant implementation details of the Umbra stealth address scheme. In Section~\ref{sec:models}, we introduce our system and threat models. In Section~\ref{sec:data_collection}, we detail our data collection methods. We present empirical Umbra usage statistics in Section~\ref{sec:umbra_activity}. In Section~\ref{sec:heuristics}, we define several heuristics aimed at deanonymizing Umbra recipients. We evaluate the efficacy of these heuristics in Section~\ref{sec:evaluation}. Finally, we conclude our paper in Section~\ref{sec:conclusion}.
\section{Background}\label{sec:background}
Hereby, we describe the stealth address protocol used by Umbra.

\subsection{Notations}\label{sec:notations}
When we uniformly at random sample $r$ from a set $S$, we write $r\in_{R}S$. We assume the existence of a prime-order cyclic group $\mathbb{G}$ ($\vert \mathbb{G}\vert = p$), where the discrete logarithm and the (decisional and computational) Diffie-Hellman assumptions hold. The finite group $\mathbb{G}$ is generated by $G$, i.e., $\mathbb{G}=\langle G\rangle$. In our applications, $\mathbb{G}$ is an elliptic curve group over a finite field $\mathbb{F}_q$ in which we denote the group operation additively. Secret keys are typically sampled from $\mathbb{F}^{*}_q$, while public keys are elliptic curve points in $\mathbb{G}$. A secure cryptographic hash function is denoted as $H(\cdot)$. Ethereum addresses are obtained by hashing the corresponding public key. Ethereum addresses are typed in $\mathsf{bold}$; sender, recipient, registrant, and stealth addresses are denoted as $\sender,\recipient,\registrant,\stealth$, respectively.
\subsection{Stealth Addresses}\label{sec:backgroundstealthaddress}
\begin{figure}[h]
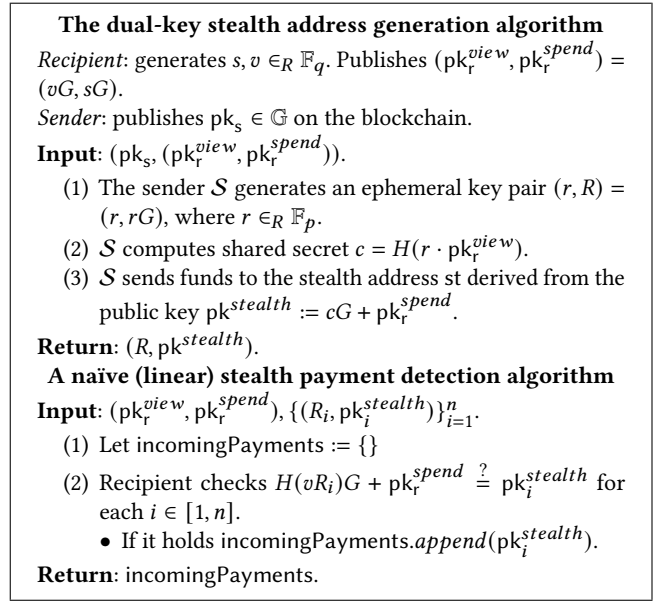

\begin{mdframed}
\centerline{\textbf{The dual-key stealth address generation algorithm}}
\textit{Recipient}: generates $s,v\in_{R}\mathbb{F}_q$. Publishes $(\pubkey^{view}_{\recipient},\pubkey^{spend}_{\recipient})=(vG,sG)$.

\textit{Sender}: publishes $\pubkey_{\sender}\in\mathbb{G}$ on the blockchain.

\textbf{Input}: $(\pubkey_{\sender},(\pubkey^{view}_{\recipient},\pubkey^{spend}_{\recipient}))$.
\begin{enumerate}
    \item The sender $\mathcal{S}$ generates an ephemeral key pair $(r,R)=(r,rG),$ where $r\in_{R}\mathbb{F}_p$.
    \item $\mathcal{S}$ computes shared secret $c=H(r\cdot\pubkey^{view}_{\recipient})$.
    \item $\mathcal{S}$ sends funds to the stealth address $\stealth$ derived from the public key $\pubkey^{stealth}:=cG+\pubkey^{spend}_{\recipient}$.
\end{enumerate}
\textbf{Return}: $(R,\pubkey^{stealth})$.

\centerline{\textbf{A naïve (linear) stealth payment detection algorithm}}

\textbf{Input}: $(\pubkey^{view}_{\recipient},\pubkey^{spend}_{\recipient}),\{(R_i,\pubkey^{stealth}_i)\}^{n}_{i=1}$.
\begin{enumerate}
    \item Let $\mathsf{incomingPayments}:=\{\}$
    \item Recipient checks $H(vR_i)G+\pubkey^{spend}_{\recipient}\stackrel{?}{=}\pubkey^{stealth}_i$ for each $i\in[1,n]$.
    \begin{itemize}
        \item If it holds $\mathsf{incomingPayments}.append(\pubkey^{stealth}_i)$.
    \end{itemize}
\end{enumerate}
\textbf{Return}: $\mathsf{incomingPayments}$.
\end{mdframed}
\caption{The dual-key stealth address protocol.}
\label{fig:stealtaddressprotocol}
\end{figure}

Umbra applies the dual-key stealth address protocol; see Figure~\ref{fig:stealtaddressprotocol}. In the applied stealth address scheme, the recipients use two key pairs: a spending and a viewing public/private key pair. The benefit of this approach is that recipients can separate the concerns of detecting incoming stealth payments and spending them. For instance, this allows recipients to accept stealth payments while keeping their spending private key off-chain, e.g., in cold storage. Additionally, this architecture enables recipients to outsource the detection of stealth payments to a trusted party since the detection of payments only relies on the secret viewing key and the spending public key. However, the trusted third party cannot spend the funds, as one also needs the spending private key to redeem the funds at the stealth address. On the other hand, this scanning approach does not provide privacy against the trusted third party.

Recently, several novel cryptographic schemes were proposed to improve the linear scanning problem of stealth addresses~\cite{beck2021fuzzy,madathil2021private,liu2022oblivious}. Nonetheless, this problem is still not entirely settled, as current solutions are unsatisfactory. Specifically, Fuzzy Message Detection~\cite{beck2021fuzzy} does not provide sufficient levels of privacy~\cite{seres2022effect}. Private Signaling assumes non-colluding scanning servers or trusted execution environments~\cite{madathil2021private,jakkamsetti2023scalable}. Both of them are strong assumptions. Oblivious Message Retrieval (OMR) and its latest variant apply fully homomorphic encryption in an elegant way~\cite{liu2022oblivious,liu2023group}. Therefore, OMR uses large ($\approx1$GB) detection keys that are impractical for resource-constrained devices. At the time of writing, OMR is being developed and soon to be deployed by the Zcash Foundation.
\section{Umbra: System and Threat models}\label{sec:models}
\subsection{Umbra: a system model}\label{sec:systemmodel}
Umbra, Ethereum's most popular stealth address scheme, consists of the following five system components.

\begin{figure}[t!]
    \centering
    \includegraphics[scale=0.5,trim={4cm 7cm 3cm 1.8cm},clip]{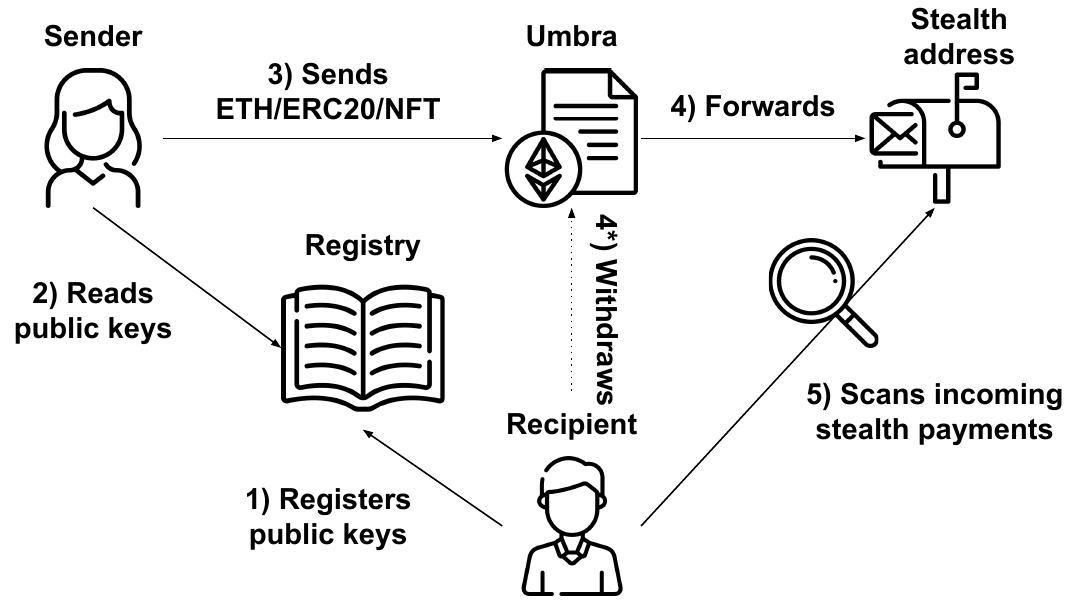}
    \caption{A schematic depiction of the Umbra stealth address scheme. A recipient first registers its viewing and spending public keys in the Stealth Key Registry. Later, the sender can read these public keys from the registry and send assets to a freshly generated pseudorandom stealth address. Finally, the recipient scans the blockchain for any incoming stealth payments and subsequently withdraws the received assets to an address they own.}
    \label{fig:umbra_schematic_description}
\end{figure}

\begin{enumerate}
    \item \textbf{Senders}: look up viewing and spending keys $(\pubkey^{view}_{\recipient},\pubkey^{spend}_{\recipient})$ of their recipients $\recipient$ in a registry and generate stealth addresses for their recipients, see Figure~\ref{fig:stealtaddressprotocol}. Subsequently, senders transfer crypto assets, e.g., ETH, or ERC20 tokens, to the stealth address for their recipients.
    \item \textbf{Recipients}: write their viewing and spending keys to a registry contract. Later, occasionally, they scan the blockchain for incoming stealth payments. Currently, they need to check every stealth address to see whether it belongs to them. Alternatively, recipients can outsource this computation. 
    \item \textbf{Registry of stealth public keys}: it is a smart contract that stores all the recipients' viewing and spending keys.
    \item \textbf{Umbra smart contract}: facilitates the asset transfer between sender and recipient. Additionally, for every stealth payment, it emits a so-called \textit{Announcement} event containing $(R,\pubkey^{stealth})$, cf. Figure~\ref{fig:stealtaddressprotocol}. Later these events can be scanned by recipients to detect incoming stealth payments.
    \item \textbf{Relayers}: in the special cases of ERC20 tokens, relayers are paid to withdraw assets from the stealth addresses. This is because those stealth addresses do not hold ether, hence cannot send transactions as they cannot pay for the occurring withdrawal transaction fees.
\end{enumerate}

The Umbra stealth address scheme entails the following steps.
\begin{enumerate}
    \item \textbf{Recipient registers public keys}: recipients register their stealth public viewing and spending keys from an Ethereum address $\registrant$ in the registry smart contract.
    \item \textbf{Sender reads registry}: senders read viewing and spending public keys $(\pubkey^{view}_{\recipient},\pubkey^{spend}_{\recipient})$ of their recipients in the stealth public key registry contract. 
    \item \textbf{Sender transfers assets}: senders generate a pseudorandom stealth address $\stealth$ given the viewing and spending public keys $(\pubkey^{view}_{\recipient},\pubkey^{spend}_{\recipient})$ and subsequently transfer assets to the derived stealth address $\stealth$.      
    \item \textbf{Recipient scans the blockchain}: recipients intermittently scan the \textit{Announcement} events emitted by the Umbra contract for incoming stealth payments. Typically, users do not leave assets residing at stealth address $\stealth$, rather, they withdraw them to a recipient address $\recipient$.

    \item \textbf{Recipient withdraws assets}: either the Umbra smart contract forwards assets to the stealth address or stores assets in the case of ERC20 tokens. In the latter case, recipients use relayers (not depicted in Figure~\ref{fig:umbra_schematic_description}) to withdraw funds held by the Umbra smart contract. Finally, recipients withdraw funds from the stealth address $\stealth$ to the recipient address $\recipient$.
\end{enumerate}
    
\subsection{Threat model}\label{sec:threatmodel}
The adversary aims to link the Umbra stealth payment recipients $\recipient$ to entities previously registered from some address $\registrant$ in the Registry contract. We assume the adversary can access the Ethereum blockchain, recording all transactions and smart contracts. Specifically, the adversary can read the code of the Umbra and Registry smart contracts and all transactions that call these smart contracts. However, we do not assume that the adversary can deanonymize Umbra users off-chain, e.g., by running a relayer node. A potent adversary on the peer-to-peer layer could certainly link IP addresses or other identifying information to Umbra users whenever they withdraw their assets from their stealth addresses. Nonetheless, our adversary solely relies on the public blockchain data to decrease the recipient anonymity and unlinkability guarantees of Umbra.

\subsection{Anonymity notions}
Stealth address schemes provide two types of privacy notions: recipient anonymity and recipient unlinkability. Informally, recipient anonymity guarantees that an observer cannot uncover the identity of a recipient better than randomly guessing. A weaker notion is recipient unlinkability, which dictates that it must be indistinguishable whether or not two stealth payments are sent to the same user. For a formal, game-based definition of these privacy notions, we refer the reader to~\cite{backes2013anoa}. Hereby, due to space constraints, we only provide a definition for recipient unlinkability; recipient anonymity can be defined analogously.

\begin{figure}
    \centering
 \begin{mdframed}
\centerline{\bf The recipient unlinkability $\mathcal{G}^{RU}_{\mathcal{A},\Pi}(\lambda)$ game}

{
\begin{enumerate}
    \item Adversary $\mathcal{A}$ selects target recipients $\recipient_0, \recipient_1$ and a target sender $\sender$.
    \item Challenger $\mathcal{C}$ instructs sender $\sender$ to send a stealth payment to $\recipient_c$ for $c\xleftarrow{\$}\{0,1\}$.
    \item $\mathcal{C}$ generates randomly a challenge bit $b\xleftarrow{\$}\{0,1\}$. If $b=0$, $\mathcal{C}$ instructs $\sender$ to send a stealth payment to $\recipient_c$. Otherwise, instructs $\sender$ to send a stealth payment to $\recipient_{1-c}$.   
    \item $\mathcal{A}$ observes the blockchain and outputs $b'$.
\end{enumerate}
\textbf{Return}: $1$, iff. $b=b'$, otherwise $0$.
}
\end{mdframed}
    \caption{The security game for the anonymity notion of recipient unlinkability.}
    \label{fig:rugame}
\end{figure}

\begin{definition}[Recipient unlinkability (RU)]
An anonymous communication protocol $\Pi$ satisfies recipient unlinkability if for all PPT adversaries $\mathcal{A}$ there is a negligible function $\mathsf{negl}(\cdot)$ such that
\begin{equation}
    \Pr[\mathcal{G}^{RU}_{\mathcal{A},\Pi}(\lambda)=1]\leq\frac{1}{2}+\mathsf{negl}(\lambda),
\end{equation}
where the privacy game $\mathcal{G}^{RU}_{\mathcal{A},\Pi}(\lambda)$ is defined in Figure~\ref{fig:rugame}.
\end{definition}

\section{Data collection}\label{sec:data_collection}
We were primarily interested in three main types of transactions in relation to the Umbra stealth address scheme on a given network.
\begin{itemize}
    \item \textbf{Stealth key registrations.} We collected every transaction that registered public keys in the stealth key registry.
    \item \textbf{Umbra sends and withdraws.} Every incoming (sent stealth payments) and outgoing (withdrawn stealth payments) transactions of the Umbra smart contract had been collected.
    \item \textbf{Registrant, sender, and withdrawer transactions.} For every address that interacted with either the Stealth Key Registry or the Umbra smart contract, we collected every transaction that was sent or received by these addresses. 
\end{itemize}
We obtained these transactions using the APIs of the Etherscan blockchain explorers for the investigated four blockchains, i.e., Ethereum L1~\footnote{See:~\url{https://etherscan.io/}}, Arbitrum~\footnote{See:~\url{https://arbiscan.io/}}, Polygon~\footnote{See:~\url{https://polygonscan.com/}}, and Optimism~\footnote{See:~\url{https://optimistic.etherscan.io/}}. Additionally, the measurements of Section~\ref{sec:umbra_activity} were obtained from the \url{https://dune.com} blockchain analytics website. 
\section{Umbra activity}\label{sec:umbra_activity}
\begin{figure}[h!]
\centering
\includegraphics[scale=0.55]{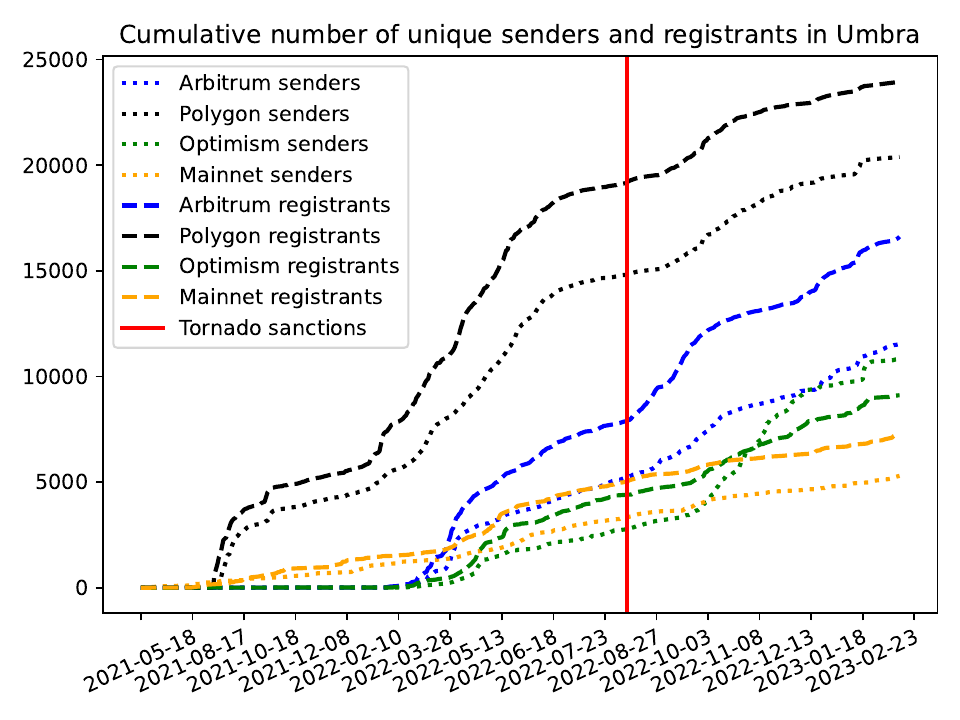}
\caption{Increasing popularity of the Umbra Stealth Address scheme on Ethereum and various Layer-2 systems. Notably, Layer-2 deployments of Umbra surpass the mainnet deployment in demand. The red vertical line shows the date when Tornado Cash contracts were sanctioned by OFAC.}
\label{fig:cumulativeSendersAndRegistrants}
\end{figure}

At the time of writing, Umbra facilitated the transfer of more than $155$ million U.S. dollars worth of cryptoassets (ether and ERC20s) across all chains where it is deployed\footnote{See:~\url{https://dune.com/intake/umbra-protocol}.}. Interestingly, Umbra became more popular on Layer 2s, such as Arbitrum, Optimism, or Polygon, than it is on mainnet Ethereum, see Figure~\ref{fig:cumulativeSendersAndRegistrants}. We observe a sharp increase in Umbra usage after the previously most popular on-chain privacy-enhancing technology, i.e., Tornado Cash, was banned in the U.S. As of July 2023, there are nearly $70,000$ registered users in the stealth key registries of Umbra across all chains.

\begin{figure}[h]
    \centering \includegraphics[scale=0.30,trim={2.5cm 1cm 0cm 2cm},clip]{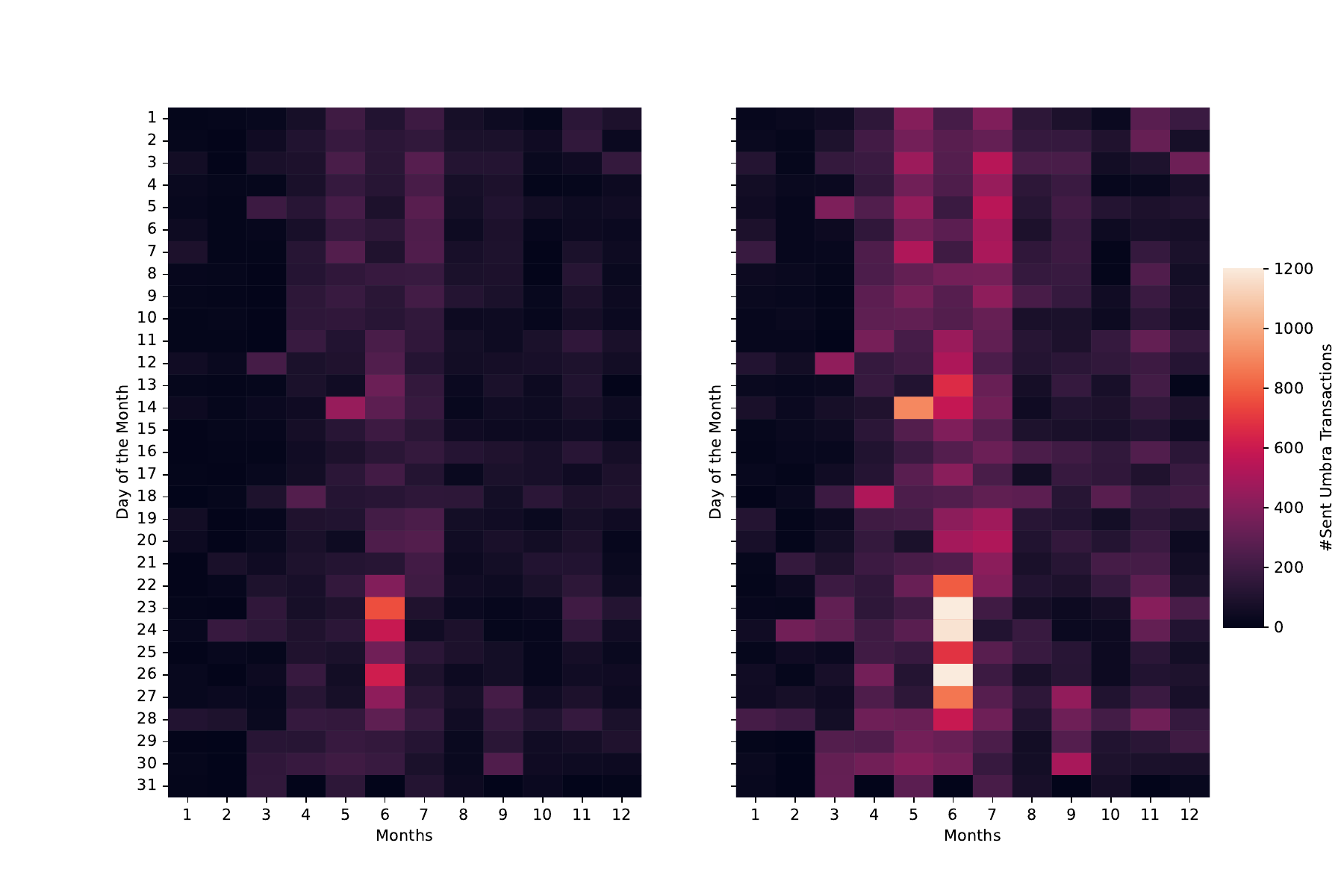}
    \caption{Umbra activity heatmaps for two Umbra users on the Polygon network. Observe that most of their activity is concentrated in a short time period. The user on the left sent $650$ Umbra transactions, while the user on the right sent $34,081$ Umbra transactions.}
    \label{fig:polygonActivity}
\end{figure}

We analyzed the usage patterns of Umbra users. First, we had suspected that certain users might use Umbra as a recipient anonymous payment processor. This could have been justified if, for instance, there are Umbra payments occurring in large numbers on a specific day of each month. Interestingly, we could not find such behavior on any of the deployed chains. See two typical activity heatmaps of Umbra users in Figure~\ref{fig:polygonActivity}. We found that numerous Umbra sender addresses are very active (i.e., send dozens of Umbra transactions) in a short time period (e.g., weeks or a few days) and afterward they do not interact with Umbra anymore. 

\section{Uncovering Umbra payment recipients}\label{sec:heuristics}
Here, we define our heuristics to identify Umbra payment recipients. We remark that our ultimate goal is to link stealth addresses to their actual recipients, i.e., addresses registered to the Umbra stealth key registry. Additionally, we provide heuristics that only reduce the anonymity guarantees of Umbra but do not uncover the actual recipient of an Umbra stealth payment deterministically.

\subsection{Registrant address reuse}\label{sec:heuristic1}
Users typically withdraw their funds (Ether or ERC20 tokens) from stealth address $\stealth$ to recipient address $\recipient$. 
If the recipient address $\recipient$ is an address already interacted with the stealth key registry, then we can heuristically link the stealth address $\stealth$ to the registrant address $\registrant$. In some cases, we can also determine the ENS (Ethereum Name Registry) address(es) owned by the recipient since, in the early days of Umbra, only ENS users could register as recipients.

\begin{heuristic}\label{heuristic1}
If a recipient address $\recipient$ is also a registrant address $\registrant$, we link the corresponding stealth address $\stealth$ to $\registrant$.
\end{heuristic}

To refine this heuristic, we only consider recipient addresses $\recipient$ where the whole amount was transferred from the stealth address $\stealth$. If this is not the case, i.e., multiple transactions originate from the stealth address potentially with different recipients, then we cannot confidently tell which recipient address (if at all) is the same entity as the owner of the stealth address. Note, we only need to check multiple outgoing withdraw transactions in the case of ether stealth transactions because, for ERC20 tokens, Umbra dictates the withdrawal of the entire stealth payment amount.

\begin{figure}
    \centering
    \includegraphics[scale=0.35,trim={0.5cm 8.5cm 0cm 2cm},clip]{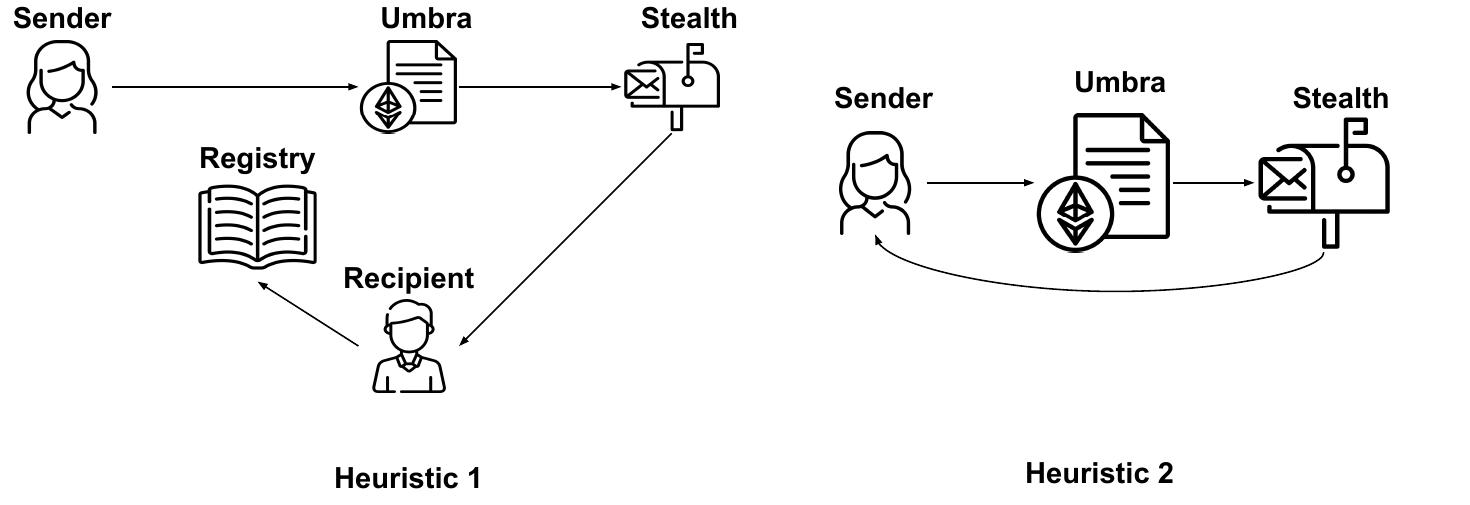}
    \caption{A schematic depiction of Heuristic~\ref{heuristic1} (Registrant address reuse), and Heuristic~\ref{heuristic2} (Same sender and receiver).}
    \label{fig:heuristics_explainer_12}
\end{figure}

\subsection{Same sender and receiver}\label{sec:heuristic2}
When the sender $\sender$ of a stealth payment is the same as the recipient $\recipient$ to where funds are withdrawn, then we heuristically link the corresponding stealth address $\stealth$ as being the same entity as the $\sender$ (and hence the $\recipient$) address. We speculate that these stealth payments are primarily issued for testing purposes.

\begin{heuristic}\label{heuristic2}
   If the sender $\sender$ of a stealth payment and the recipient address $\recipient$ of a withdraw transaction are the same, then we link the stealth address $\stealth$ and $\sender$ as being the same entity.  
\end{heuristic}

Similarly, in this heuristic, we check whether the whole amount was withdrawn from the stealth address $\stealth$ to $\recipient$. If yes, then we apply the heuristic and conclude that $\sender$, $\stealth$, and $\recipient$ addresses are owned by the same entity. Otherwise, we do not apply the heuristic. If there are multiple withdraw transactions from a stealth address with possibly multiple recipient addresses, then we cannot confidently link the actual owner of the stealth address $\stealth$ to the sender address $\sender$ even if one of the recipient addresses is the same as the $\sender$.

\begin{figure}
    \centering
    \includegraphics[scale=0.5,trim={10.5cm 9cm 0.1cm 0.2cm},clip]{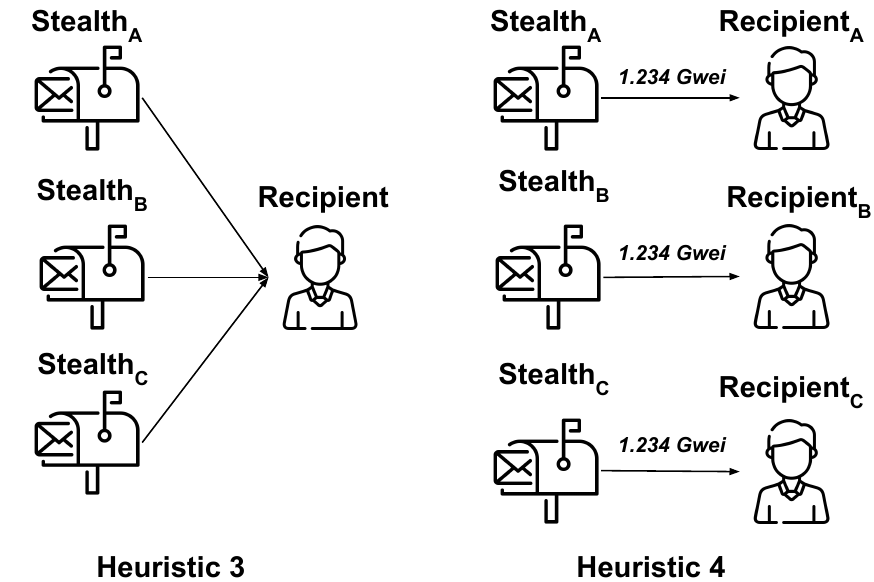}
    \caption{A schematic depiction of Heuristic~\ref{heuristic3} (Collector pattern), and Heuristic~\ref{heuristic4} (Unique \texttt{maxPriorityFeePerGas}). Both of these heuristics forfeit the unlinkability guarantee provided by Umbra, as these heuristics enable one to link the recipients of multiple stealth payments.}
    \label{fig:heuristics_explainer_34}
\end{figure}

\subsection{Collector pattern}\label{sec:heuristic3}
A stealth address scheme should provide recipient anonymity and \textit{recipient unlinkability}. Recipient unlinkability dictates that the adversary should not be able to tell whether or not two messages (or payments) are sent to the same user. An anonymous communication system satisfies recipient unlinkability if the adversary cannot do better than randomly guessing whether two messages are sent to the same user~\cite{backes2013anoa}. The following heuristic aims to break the recipient unlinkability guarantees of Umbra.

Umbra users can forfeit the recipient unlinkability guarantees the stealth address scheme provides.
If a set of stealth addresses $\{\stealth_i\}^{n}_{i=0}$ have sent their funds directly to the same recipient address $\recipient$, then we can cluster the stealth addresses $\{\stealth_i\}^{n}_{i=0}$ as they are likely owned by the same entity. This implies that the recipients of these stealth payments are likely to be the same entity. Note that this heuristic is not able to link a recipient $\recipient$ or stealth address $\stealth$ to a registrant address $\registrant$ in the Umbra stealth key registry. Nonetheless, it already uncovers important information about a set of stealth addresses.

\begin{heuristic}\label{heuristic3}
   If a user withdraws funds from a set of stealth addresses $\{\stealth_i\}^{n}_{i=0}$ to the very same recipient address $\recipient$, then we cluster all these addresses ($\{\stealth_i\}^{n}_{i=0}$ and $\recipient$) together as they are likely controlled by the same entity.
\end{heuristic}

We acknowledge the possibility of false positives output by Heuristic~\ref{heuristic3}. For instance, it might be that the recipient address $\recipient$ is a business that provides services in exchange for ether or ERC20 tokens, and Umbra users with stealth addresses are paying for them. We refine this heuristic as follows in order to avoid false positives. First, we only consider stealth addresses from where there is only a single withdrawal transaction, i.e., the whole amount is withdrawn to a recipient address in one transaction. If there were multiple transactions from a stealth address, then it is more likely that they are rather payment transactions and not withdrawal transactions. We remark that one only needs to check this condition in the case of ether transactions because, for ERC20 tokens, users must withdraw all the amount from the Umbra contract in one transaction.

\subsection{Unique max priority fee}\label{sec:heuristic4}
In contrast to the previous Heuristic~\ref{heuristic3} (collector pattern), it is possible to link stealth addresses even if a user withdraws from a stealth address $\stealth_{i}$ to a different recipient address $\recipient_{i}$ for $i\in[1,n]$.

Ethereum uses a novel transaction fee mechanism called EIP-1559~\cite{buterin2019eip} that was deployed on August 5, 2021, as part of the London hard fork update. For our discussion, the only relevant part of the EIP-1559 transaction fee mechanism design is the \texttt{maxPriorityFeePerGas} field of Ethereum transactions. This is the only part of an Ethereum transaction set by the user (wallet) concerning transaction fees. The larger the transaction fee \texttt{maxPriorityFeePerGas} is, the sooner a validator includes the transaction in a new block. Typically, users (rather their wallet softwares) set \texttt{maxPriorityFeePerGas} automatically. However, users can modify the \texttt{maxPriorityFeePerGas} field manually and leave a chosen setting for an extended period of time. Hence, if a wallet owns multiple addresses and users apply ``unique'' \texttt{maxPriorityFeePerGas} fees, then the sender addresses of these transactions can be clustered together as they are likely owned by the same entity. 

\begin{heuristic}\label{heuristic4}
   If two or more withdraw transactions from $\stealth_{i}\rightarrow\recipient_{i}$ uses the same ``unique'' \texttt{maxPriorityFeePerGas}, then we link all these stealth addresses $\{\stealth_{i}\}^{n}_{i=1}$ together. 
\end{heuristic}

To make this Heuristic sensible, we need to choose when we consider a \texttt{maxPriorityFeePerGas} value ``unique'' in our dataset. Obviously, if a value is encountered dozens of times, then we cannot consider it unique, e.g., this is the case with round number as \texttt{maxPriorityFeePerGas} values. To limit the number of false positives, we considered a \texttt{maxPriorityFeePerGas} values ``unique'' if not more than five transactions apply that \texttt{maxPriorityFeePerGas} value. Note that this Heuristic does not apply to Umbra stealth payments that transfer ERC-20 tokens. Tokens are withdrawn by relayers on behalf of Umbra recipients. Therefore, token withdrawal transactions' \texttt{maxPriorityFeePerGas} field is not set by Umbra recipients but rather by the relayers.

\section{Evaluation}\label{sec:evaluation}
In this section, we evaluate the effectiveness of our heuristics in identifying Umbra payment recipients.
\subsection{Heuristic~\ref{heuristic1}}\label{sec:eval_heuristic1}
The first heuristic is the most successful in identifying the true recipients of Umbra payments; $48,25\%$ of Umbra payments (both ether and ERC20 token transfers) on the main Ethereum network can be trivially linked to stealth key registry transactions as the withdrawer address, and the registrant addresses are the same. We observe similarly high percentages for the other networks, i.e., $25,79\%$, $65,61\%$, $52,57\%$ for Polygon, Arbitrum, and Optimism, respectively. 
\subsection{Heuristic~\ref{heuristic2}}\label{sec:eval_heuristic2}
This heuristic (same sender and withdrawer address) performs worse than the first heuristic. However, it still identifies a handful of Umbra withdraw transactions, i.e., $2,61\%$, $1,15\%$, $1,87\%$, and $0,85\%$  of Umbra payments are deanonymized on the Ethereum L1, Polygon, Arbitrum, and Optimism networks, respectively, thanks to this heuristic.
\subsection{Heuristic~\ref{heuristic3}}\label{sec:eval_heuristic3}
\begin{figure}[h!]
    \centering
\includegraphics[scale=0.55]{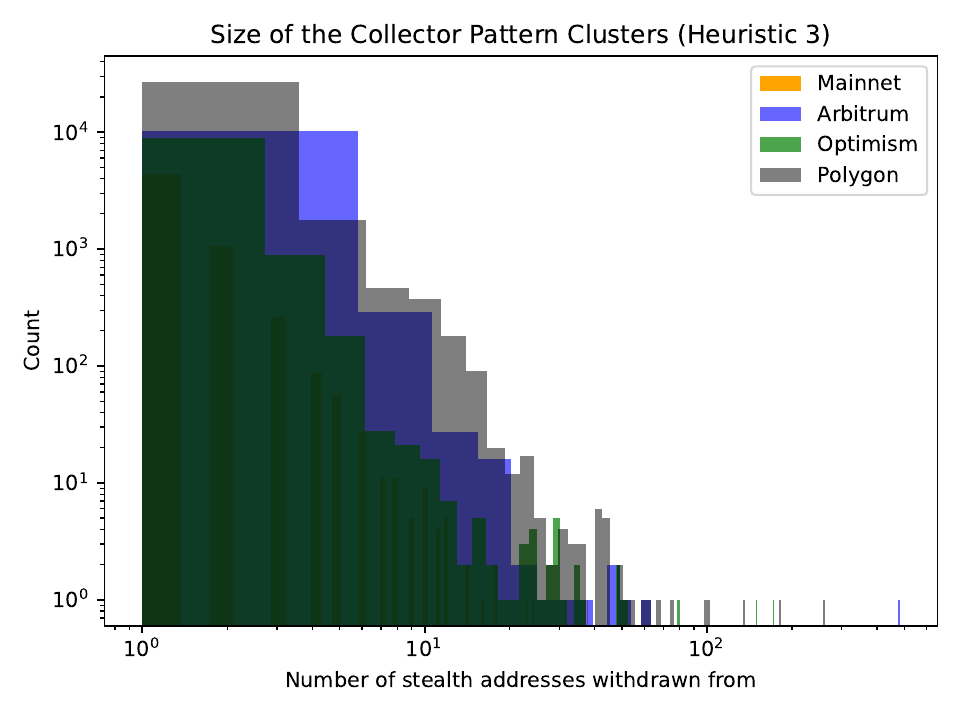}
    \caption{This figure displays the number of withdrawer addresses with a given number of withdraw transactions from stealth addresses. Users tend to reuse withdraw addresses to collect stealth payments from multiple stealth addresses. Note that this is a log-log figure.}
    \label{fig:collectorPattern}
\end{figure}

The collector pattern heuristic breaks the recipient unlinkability guarantee of Umbra payments. As Figure~\ref{fig:collectorPattern} shows, several Umbra users withdraw multiple incoming payments to a \emph{single} withdraw address. The largest whale is a user on the Arbitrum network who collected $481$ stealth payments to a unique withdrawer address.

\begin{table}[h!]
\centering
 \begin{tabular}{l c c c c} 
 \hline
  & \textbf{Mainnet} & \textbf{Polygon} & \textbf{Arbitrum} & \textbf{Optimism} \\ [0.5ex] 
 \hline
 Heuristic~\ref{heuristic3} & $8.48$\ bits & $9.86$\ bits & $8.86$\ bits & $8.83$\ bits \\ 
Entropy  & $12.53$\ bits & $14.86$\ bits & $13.36$\ bits  & $13.29$\ bits \\[1ex] 
 \hline
 \end{tabular}
 \caption{Most users withdraw multiple incoming stealth payments to a single address. This user behavior, formulated as ``collector pattern'' in Heuristic~\ref{sec:heuristic3}, greatly reduces the Shannon entropy of stealth payment recipients.}
 \label{table:heuristics3EntropyEvaluation}
\end{table}

If $n$ stealth payments are occurring on a network, an adversary ideally has $\log_2(n)$ bits of uncertainty about the recipients of stealth payments. However, with the help of the ``collector pattern'' heuristic, we can heuristically conclude that several stealth payments are likely received by the same user, hence, decreasing the entropy of this distribution, see Table~\ref{table:heuristics3EntropyEvaluation}. For instance, in the case of Arbitrum, naïvely, there is a $13,36$ bits of entropy in the recipients' distribution. The application of Heuristic~\ref{heuristic3} reduces this to $8.86$ bits.

\subsection{Heuristic~\ref{heuristic4}}\label{sec:eval_heuristic4}
Previous work had already successfully applied the heuristic of unique transaction fees in the context of linking Tornado cash deposit and withdraw transactions~\cite{beres2021blockchain}. Interestingly, in our case, we did not find Umbra withdraw transactions with the same unique priority fees. We attribute this phenomena as one of the beneficial outcomes of the EIP-1559 transaction fee mechanism.

\begin{table}[h!]
\centering
 \begin{tabular}{l c c c c} 
 \hline
  & \textbf{Mainnet} & \textbf{Polygon} & \textbf{Arbitrum} & \textbf{Optimism} \\ [0.5ex] 
 \hline
 Heuristic~\ref{heuristic1}. & $4671$ & $15075$ & $12488$ & $8391$ \\ 
 Heuristic~\ref{heuristic2}.  & $253$ & $670$ & $356$  & $135$\\
 \hline
 Total linked   & $4696$ & $15084$ & $12513$ & $8403$ \\
 Total withdrawn   & $9680$ & $58454$ & $19033$ & $15963$ \\[1ex] 
 \hline
 \end{tabular}
 \caption{Performance of the identified heuristics for uncovering Umbra payment recipients. Three of our heuristics could link the $48.5\%$, $25.8\%$, $65.7\%$, and $52.6\%$ of all Umbra transactions on the Ethereum L1, Polygon, Arbitrum, and Optimism networks, respectively.}
 \label{table:heuristicsEvaluation}
\end{table}

\subsection{Countermeasures}\label{sec:countermeasures}
A common root cause of Heuristic~\ref{sec:heuristic1},~\ref{sec:heuristic2}, and~\ref{sec:heuristic3} is the ability to reuse user addresses. Generally speaking, one should always avoid the reuse of addresses even in an account-based cryptocurrency as it facilitates user profiling and address clustering~\cite{moser2022resurrecting}. The use of unique addresses is especially important when one uses privacy-enhancing technologies~\cite{beres2021blockchain,wang2023zero,wu2022tutela}. One must use different addresses for different types of on-chain activities. Developers of stealth address wallets should never allow users to reuse addresses or to withdraw funds from a stealth address to a registrant address.   
\section{Conclusion and future directions}\label{sec:conclusion}
In this work, we analyzed the recipient anonymity and unlinkability guarantees provided by the second most used privacy-enhancing overlay in the Ethereum DeFi ecosystem. We identified four heuristics that enabled us to link the majority of Umbra recipients on the Ethereum mainnet. Finally, we suggested several countermeasures that would prevent one from our identified heuristics. Future work should investigate the interplay between Ethereum's cross-chain ecosystem and the deployed privacy-enhancing technologies on these chains. For instance, Ethereum and its L2s use the same address format. We also observed that users tend to reuse their addresses across different networks. We suspect that this address reuse could lead to cross-chain heuristics that could further decrease the recipient anonymity guarantees of Umbra. Another venue for future work could be to apply the heuristics of~\cite{victor2020address} to enhance our deanonymization and linking capabilities.


\begin{acks}
We are grateful to Ben DiFrancesco, Matt Solomon, and Soptq for insightful discussions.
\end{acks}

\bibliographystyle{ACM-Reference-Format}
\bibliography{sample-base}

\appendix

\end{document}